
\input epsf

\ifx\epsffile\undefined\message{(FIGURES WILL BE IGNORED)}
\def\insertfig#1#2{}
\else\message{(FIGURES WILL BE INCLUDED)}
\def\insertfig#1#2{{{\baselineskip=4pt
\midinsert\centerline{\epsfxsize=\hsize\epsffile{#2}}{{
\centerline{#1}}}\medskip\endinsert}}}
\fi
\input harvmac

%
%
%
%
%
\ifx\answ\bigans
\else
\output={

\almostshipout{\leftline{\vbox{\pagebody\makefootline}}}\advancepageno

}
\fi
%
%
%

%
%

%
%
\def\UCSD#1#2{\noindent#1\hfill #2%
\bigskip\supereject\global\hsize=\hsbody%
\footline={\hss\tenrm\folio\hss}}
%
%
\def\abstract#1{\centerline{\bf Abstract}\nobreak\medskip\nobreak\par
#1}
%
%
%
%
\edef\tfontsize{ scaled\magstep3}
 \tfontsize  \tfontsize
\font\titlermss=cmr5 \tfontsize \font\titlei=cmmi10 \tfontsize
\font\titleis=cmmi7 \tfontsize \font\titleiss=cmmi5 \tfontsize
\font\titlesy=cmsy10 \tfontsize \font\titlesys=cmsy7 \tfontsize
\font\titlesyss=cmsy5 \tfontsize  \tfontsize
\skewchar\titlei='177 \skewchar\titleis='177 \skewchar\titleiss='177
\skewchar\titlesy='60 \skewchar\titlesys='60 \skewchar\titlesyss='60
\scriptscriptfont0=\titlermss
\scriptscriptfont1=\titleiss
\scriptscriptfont2=\titlesyss
%
%
%

%
\def\inv{^{\raise.15ex\hbox{${\scriptscriptstyle -}$}\kern-.05em 1}}
\def\lbar{{\lower.35ex\hbox{$\mathchar'26$}\mkern-10mu\lambda}}

%
%
%
%
\def\slash#1{\rlap{$#1$}/} 
\def\dsl{\,\raise.15ex\hbox{/}\mkern-13.5mu D} 
\def\delsl{\raise.15ex\hbox{/}\kern-.57em\partial}
\def\Ksl{\hbox{/\kern-.6000em\rm K}}
\def\Asl{\hbox{/\kern-.6500em \rm A}}
\def\Dsl{\hbox{/\kern-.6000em\rm D}} 
\def\Qsl{\hbox{/\kern-.6000em\rm Q}}
\def\gradsl{\hbox{/\kern-.6500em$\nabla$}}
%
%
\def\lspace{\ifx\answ\bigans{}\else\qquad\fi}
\def\lbspace{\ifx\answ\bigans{}\else\hskip-.2in\fi} 
%
%
\def\boxeqn#1{\vcenter{\vbox{\hrule\hbox{\vrule\kern3pt\vbox{\kern3pt
        \hbox{${\displaystyle #1}$}\kern3pt}\kern3pt\vrule}\hrule}}}
%
%
\def\mbox#1#2{\vcenter{\hrule \hbox{\vrule height#2in
\kern#1in \vrule} \hrule}}
%
%
%
%

%
%
%
%
%

%

\def\bar#1{\overline{#1}}

\def\darr#1{\raise1.5ex\hbox{$\leftrightarrow$}\mkern-16.5mu #1}

%
%
\def\frac#1#2{{\textstyle{#1\over #2}}} 
%
%
%
%

%
%
%
%

%
%
\def\ltap{\ \raise.3ex\hbox{$<$\kern-.75em\lower1ex\hbox{$\sim$}}\ }
\def\gtap{\ \raise.3ex\hbox{$>$\kern-.75em\lower1ex\hbox{$\sim$}}\ }
\def\gl{\ \raise.5ex\hbox{$>$}\kern-.8em\lower.5ex\hbox{$<$}\ }
\def\roughly#1{\raise.3ex\hbox{$#1$\kern-.75em\lower1ex\hbox{$\sim$}}}

%
%

%

%
\def\np#1#2#3{{Nucl. Phys. } B{#1} (#2) #3}
\def\pl#1#2#3{{Phys. Lett. } {#1}B (#2) #3}
\def\prl#1#2#3{{Phys. Rev. Lett. } {#1} (#2) #3}
\def\physrev#1#2#3{{Phys. Rev. } {#1} (#2) #3}

\relax

\def\hbar{\bar h_Q}

\def\qsl{\hbox{/\kern-.5600em {$q$}}}
\def\ksl{\hbox{/\kern-.5600em {$k$}}}

\def\({\left(}
\def\){\right)}

\def\OMIT#1{}
\def\frac#1#2{{#1\over#2}}

\def\semi{$\Lambda_b\rightarrow\Lambda_ce^-\overline{\nu}_e$}

\hbadness=10000

\noblackbox
\vskip 1.in
\centerline{{\titlefont{Long-Distance Contributions to the }}}
\medskip
\centerline{{\titlefont{$\Lambda_b\rightarrow\Lambda_ce^-\overline{\nu}_e$
Isgur-Wise Function}}}
\medskip
\vskip .5in
\centerline{Martin J.~Savage}
\medskip
{\it{ \centerline{ Department of Physics, Carnegie Mellon University,
Pittsburgh PA 15213} }}

\vskip .5in

\abstract{ The six independent formfactors describing the weak decay  \semi\
are
determined in terms of one universal function when both the $b$ and $c$
quarks are treated as heavy.
The relations between these formfactors are modified by terms
suppressed by inverse powers of the charm quark mass.
We compute the formally leading long-distance corrections to these relations
in chiral perturbation theory both at and away from zero recoil.
They arise  from the spin-symmetry breaking
$\Sigma_c^* - \Sigma_c$ mass splitting and are nonanalytic in the pion mass.
The correction at the zero-recoil point is found to be
$ - (3.4 \pm 2.0) \times 10^{-4} g_3^2$
for a quark model estimate of $m_{\Sigma_c^*}-m_{\Sigma_c}$ and
where $g_3\sim 1$ is an axial coupling constant.
Further, the formfactors that vanish in the heavy quark limit do not receive
such corrections.  }

\vfill
\UCSD{\vbox{
\hbox{CMU-HEP 94-04}
\hbox{DOE-ER/40682-58}}
}{January  1994}
\eject

Great simplifications occur in the strong dynamics of a quark in the limit that
its
mass becomes infinitely larger than the QCD scale
\ref\iwhq{N. Isgur and M.B.Wise, \pl{232}{1989}{113};
\pl{237}{1990}{527}.}\nref\ehhq{E. Eichten and B. Hill,
\pl{234}{1990}{511}.}-\ref\gehq{H. Georgi, \pl{240}{1990}{447}.} .
The new spin-flavour symmetries that become manifest in this limit of QCD
give rise to stringent constraints on weak matrix elements between hadrons
containing heavy quarks.
Apriori, six independent formfactors are required to describe the decay \semi\
{}.
When both the $b$ and $c$ quarks are treated as heavy these formfactors are all
related
in terms of one universal function
\ref\IWbary{N. Isgur and M.B. Wise, \np{348}{1991}{276}.}
\ref\geogib{H. Georgi, \np{348}{1991}{293}.}.
The vanishing of the leading $1/m_c$ corrections at zero-recoil to this decay
\ref\GGW{H. Georgi, B. Grinstein and M.B. Wise, \pl{252}{1990}{456}.}
provides another opportunity for a model independent measurement of the
weak mixing angle  $V_{bc}$.
In this work we will compute the formally leading corrections to the relations
between the six formfactors arising from long-distance physics.
These corrections are nonanalytic in the pion mass
and arise from the spin-symmetry breaking  $\Sigma_c^* - \Sigma_c$ mass
splitting.
Analogous computations have been performed in the meson sector
\ref\randwise{L. Randall and M.B.Wise, \pl{303}{1993}{139}.}
\ref\chowwise{C. Chow and M.B.Wise, \physrev{D48}{1993}{5202}.}.

In general there are six independent form factors $f_i$ and $g_i$
(conventionally functions of $q^2$)
describing the decay \semi\  defined by
\eqn\form{ \eqalign{
\langle \Lambda_c (p^\prime) | \overline{c}\gamma^\mu b | \Lambda_b(p) \rangle
&  = \overline{u}_{\Lambda_c}(p^\prime)  \left[ f_1\gamma^\mu
-if_2\sigma^{\mu\nu}q_\nu
+ f_3q^\mu \right]  u_{\Lambda_b}(p)\cr
\langle \Lambda_c (p^\prime) | \overline{c}\gamma^\mu\gamma_5 b | \Lambda_b(p)
\rangle
&  = \overline{u}_{\Lambda_c} (p^\prime ) \left[ g_1\gamma^\mu
-ig_2\sigma^{\mu\nu}q_\nu
+ g_3q^\mu \right]\gamma_5  u_{\Lambda_b} (p) }\ \ \ \ \ ,}
where $q=p-p^\prime$ is the momentum transfer to the leptons.
As we wish to treat the $b$ and $c$ quark as heavy compared the the scale of
strong
interactions it is convenient to write the matrix element in terms of the
four-velocities $v$
and $v^\prime$ instead of the momentum $p$ and $p^\prime$ which diverge in the
infinite mass limit.
The formfactors $F_i$ and $G_i$
( functions of $w=v\cdot v^\prime$ for convenience) are defined by
\eqn\vec{ \eqalign{
\langle \Lambda_c (v^\prime) | \overline{c}\gamma^\mu b | \Lambda_b(v) \rangle
&  = \overline{u}_{\Lambda_c} \left[ F_1\gamma^\mu + F_2v^\mu +
F_3v^{\prime\mu}\right]
u_{\Lambda_b}\cr
\langle \Lambda_c (v^\prime) | \overline{c}\gamma^\mu\gamma_5 b | \Lambda_b(v)
\rangle
&  = \overline{u}_{\Lambda_c} \left[ G_1\gamma^\mu + G_2v^\mu +
G_3v^{\prime\mu}
\right]\gamma_5  u_{\Lambda_b} }\ \ \ \ \ .}
It is easy to show that
\eqn\rel{\eqalign{
f_1 = & F_1+(m_{\Lambda_b}+m_{\Lambda_c})({1\over 2m_{\Lambda_b}}F_2 +
{1\over 2m_{\Lambda_c}}F_3) \cr
f_2 = & -( {1\over 2m_{\Lambda_b}}F_2 + {1\over 2m_{\Lambda_c}}F_3) \cr
f_3 = & ({1\over 2m_{\Lambda_b}}F_2 - {1\over 2m_{\Lambda_c}}F_3) \cr
g_1 = & G_1-(m_{\Lambda_b}-m_{\Lambda_c})({1\over 2m_{\Lambda_b}}G_2 +
{1\over 2m_{\Lambda_c}}G_3) \cr
g_2 = & -( {1\over 2m_{\Lambda_b}}G_2 + {1\over 2m_{\Lambda_c}}G_3) \cr
g_3 = & ({1\over 2m_{\Lambda_b}}G_2 - {1\over 2m_{\Lambda_c}}G_3) \cr } \ \ \ \
.}

When the $b$ quark is treated as heavy the
formfactors are determined in terms of two independent functions
${\cal F}_1$ and  ${\cal F}_2$
\ref\manroba{T. Mannel and W. Roberts, \physrev{D47}{1993}{4963}.} .
It is the spin symmetry of the $b$ quark alone that allows the matrix element
to become
\eqn\inter{
\langle \Lambda_c (v^\prime) | \overline{c}\Gamma h^{(b)}_v | \Lambda_b(v)
\rangle
 = \overline{u}_{\Lambda_c} \left[ {\cal F}_1 + {\cal F}_2\slash{v}\right]
\Gamma
u_{\Lambda_b}  \ \ \ ,}
for any lorentz structure  $\Gamma$.
The dynamical QCD quark field $b$ has been integrated out and replaced with
a  static quark field $h^{(b)}_v$ of four-velocity $v$.
Apriori, there are no constraints on the functions ${\cal F}_1$ and ${\cal
F}_2$,
in particular, there is no normalisation condition at the zero-recoil point
$w=1$.
As only the $b$ quark has been treated as heavy there are relations between
the formfactors  that are valid for any value of the charm quark mass \manroba\
,
\eqn\bspin{\eqalign{ f_1 = & g_1 = {\cal F}_1 + {m_{\Lambda_c}\over
m_{\Lambda_b}}{\cal F}_2\cr
f_2 = & -f_3 = g_2 = -g_3 = -{1\over m_{\Lambda_b}}{\cal F}_2 } \ \ \ \ .}
Such relations will receive corrections of the form $\Lambda_{\rm QCD}/m_b$
and $\alpha_s(m_b) m_c/m_b$ (which we neglect in this work)
but will be valid to all orders in a $1/m_c$ and $\alpha_s(m_c)$ expansion.
When the $c$ quark is also treated as heavy compared to the QCD scale
it too can be integrated out and replaced by a static field
$h^{(c)}_{v^\prime}$ of four-velocity $v^\prime$.
This then allows all six formfactors to be written in terms of one
universal function, $\eta^{(\Lambda)} (w)$.
The weak matrix element becomes
\eqn\IWa{  \langle \Lambda_c (v^\prime) | \overline{h}^{(c)}_{v^\prime}
 \Gamma h^{(b)}_v | \Lambda_b(v) \rangle
 = \eta^{(\Lambda)} (w) \overline{u}_{\Lambda_c}\Gamma u_{\Lambda_b}  \ \ \ .}

The Isgur-Wise function $\eta^{(\Lambda)} (w)$ at zero-recoil ($w=1$) is
related to the $b$ number current via the flavour symmetry
in the heavy quark limit and consequently  $\eta^{(\Lambda)} (1) = 1 $.
It follows from \IWa\ that $f_1=g_1= C_{bc}\ \eta^{(\Lambda)} (w)$ and
$f_2=f_3=g_2=g_3=0$ where $C_{bc}$ arises from perturbative strong
interactions between the scale of the $b$ and $c$ quarks and is calculable
\ref\falkital{A. Falk etal., \np{343}{1990}{1}.}.
The leading $1/m_c$ corrections to these relations have been computed in \GGW\
\eqn\first{\eqalign{
f_1 = g_1 = & \  G_1\left( 1+ {\overline{\Lambda}\over 2 m_c}{1\over 1+w}
\left(1-{m_{\Lambda_c}\over m_{\Lambda_b}}\right) \right) \cr
f_2 = -f_3 = &\ g_2 = -g_3 =  {G_1\over m_{\Lambda_b}}
{\overline{\Lambda}\over 2 m_c}{1\over 1+w} }\ \ \ \ ,}
where $G_1$ is the formfactor appearing in \rel\ normalised at zero recoil to
$C_{bc}$ and $\overline{\Lambda} = m_{\Lambda_Q}-m_Q$.
Such corrections do not renormalise the matrix element at zero recoil.
These relations are modified by terms higher order in $1/m_c$
\ref\falkneubert{A.F.Falk and M. Neubert, \physrev{D47}{1993}{2982}.}
and
$\alpha_s(m_c)$ but in such a way to preserve \bspin\ ,
it is a class of these corrections we will compute in this work.

The dynamics of the pseudo-Goldstone bosons associated with the spontaneous
breaking of $SU(3)_L\otimes SU(3)_R$ chiral symmetry to $SU(3)_V$ symmetry
are described by the lagrangian
\eqn\pion{ {\cal L}_\pi = {f_\pi^2\over 8}Tr \left( \partial^\mu\Sigma^\dagger
\partial_\mu\Sigma \right) + ..........\ \ \ ,}
where the dots denote operators with more derivatives or insertions of the
light quark mass matrix.
With this definition the pion decay constant $f_\pi = 135$MeV and $\Sigma$
 is the exponential of the pseudo-Goldstone boson field,
\eqn\field{\Sigma = {\rm exp}\left( 2iM/f_\pi\right)\ \ \ \ \ ,}
where $M$ is the octet of pseudo-Goldstone bosons
\eqn\mesons{ M = \left(
\matrix{ {1\over\sqrt{6}}\eta + {1\over\sqrt{2}}\pi^0 & \pi^+ & K^+ \cr
                                                      \pi^- &
{1\over\sqrt{6}}\eta - {1\over\sqrt{2}}\pi^0 &
K^0 \cr
                                                        K^- & \overline{K}^0 &
-{2\over\sqrt{6}}\eta \cr } \right) \ \ \ \ \ .}
The pseudo-Goldstone boson field has a well defined transformation
under chiral symmetry $\Sigma\rightarrow L\Sigma R^\dagger$ however in order
to introduce matter fields such as baryons or mesons containing a heavy quark
it is convenient to introduce the field $\xi$ which is defined as $\xi^2 =
\Sigma$.
This transforms under chiral symmetry as
\eqn\xitrans{\xi\rightarrow L\xi U^\dagger =  U\xi R^\dagger  \ \ \ \ ,}
which implicitly defines $U$, a matrix that depends upon the pion field and on
$L$ and $R$.
{}From this, two vector fields of definite parity can be constructed
\eqn\vecxi{
V^\mu = {1\over 2}\left( \xi\partial^\mu\xi^\dagger +
\xi^\dagger\partial^\mu\xi \right) \ , \ \ \ \ \ \
A^\mu = {i \over 2}\left( \xi\partial^\mu\xi^\dagger  -
\xi^\dagger\partial^\mu\xi \right) \ \ \ \ \ \ \ .}
The vector field $V^\mu$ transforms inhomogeneously under chiral symmetry
$V^\mu\rightarrow UV^\mu U^\dagger + U\partial U^\dagger$ while the axial
vector field transforms homogeneously $A^\mu\rightarrow UA^\mu U^\dagger $.

It is known how to combine heavy quark symmetries and chiral symmetry together
in order to describe the soft hadronic interactions of hadrons containing a
heavy quark
\ref\mbwchir{M. B. Wise, \physrev{D45}{1992}{2188}.}\nref\burd{G. Burdman and
J. Donoghue, \pl{208}{1992}{287}.}\nref\yan{T. M. Yan etal.,
\physrev{D46}{1992}{1148}.}-\ref\cho{P. Cho, \pl{285}{1992}{145}.} .
As we are only concerned with heavy baryons we will not
discuss the lagrangian for heavy mesons, we refer the reader to
\ref\mbwlouise{M.B. Wise, Lectures given at the CCAST Symposium on Particle
Physics at the Fermi Scale, May 1993.}
for a review.
Light degrees of freedom in the ground state of a baryon containing one heavy
quark
can have $s_l=0$ corresponding to a member of the
$SU(3)_V$ antitriplet, $T_i (v)$
or they can have $s_l=1$ corresponding to a member of the $SU(3)_V$ sextet ,
$S^{ij}_\mu (v)$.
In the latter case, the spin of the light degrees of freedom can be combined
with the
spin of the heavy
quark to form both $J=3/2$ and $J=1/2$ baryons, which are degenerate in the
$m_Q\rightarrow \infty$ limit.
The sextet field $S^{ij}_\mu (v)$ (satisfying $v^\alpha S^{ij}_\alpha (v)=0$)
contains twelve
baryon fields, six $J=1/2$ and six $J=3/2$ fields.   Using the notation of
\cho\ ,
\eqn\tripsex{ \eqalign{
S^{ij}_\mu (v) = & {1\over\sqrt{3}} (\gamma_\mu + v_\mu) \gamma_5
{1\over 2} (1+\slash{v}) B^{ij}  \ + \ {1\over 2}(1+\slash{v}) B^{*ij}_\mu \cr
T_i (v) = & {1\over 2}(1+\slash{v}) B_i } \ \ \ \ ,}
where the elements of the symmetric tensor $B^{ij}$ containing
the charm baryons are
\eqn\sex{\eqalign{
B^{11} = \Sigma_c^{++}\ ,\ B^{12} = {1\over\sqrt{2}}\Sigma_c^+\ ,\ B^{22} =
\Sigma_c^0 \ ,\cr
B^{13} = {1\over\sqrt{2}}\Xi_{c2}^+ \ ,\ B^{23} = {1\over\sqrt{2}}\Xi_{c2}^0 \
,\
B^{33} = \Omega_c^0 }\ \ \ .}
The $J=3/2$ partners of these baryons (described by the Rarita-Schwinger
field $B^{*ij}_\mu$ satisfying $\gamma^\mu B^{*ij}_\mu = 0$)  have the same
$SU(3)_V$ assignment.
The elements of the antitriplet representation are for the charm baryons
\eqn\trip{B_1 = \Xi_{c1}^0 \  , \ B_2 = -\Xi_{c1}^+ \ , \ B_3 = \Lambda_c^+ \ \
\ .}

The chiral lagrangian describing the soft hadronic interaction of these baryons
is given by \cho\
\eqn\lag{ \eqalign{
{\cal L}_Q = &\ \  i\overline{T^i} v\cdot D T_i \ -\  i\overline{S}^\mu_{ij}
v\cdot D S_\mu^{ij}
\ +\ \Delta_0 \overline{S}^\mu_{ij} S_\mu^{ij}\cr
&\ + \ g_3\left(  \epsilon_{ijk}\overline{T}^i (A^\mu)^j_l S^{kl}_\mu + h.c.
\right)
+  \ i g_2\ \epsilon_{\mu\nu\rho\sigma}\overline{S}^\mu_{ik} v^\nu(A^\rho)^i_j
S^{\sigma jk} + ............}  \ \ ,}
where the dots denote operators with more insertions of the light quark mass
matrix,
more derivatives or higher order in the $1/m_Q$ expansion and
$D^\alpha$ is the chiral covariant derivative.
Coupling of the pseudo-Goldstone bosons to the antitriplet baryons
is forbidden at lowest order in $1/m_Q$.
Even in the infinite mass limit the $\Sigma_Q^{(*)}$ baryons are not degenerate
with the $\Lambda_Q$ baryons as the light degrees of freedom are
in a different configuration.  This intrinsic mass difference is $\Delta_0$.

The left handed weak current responsible for the $b\rightarrow c$ transition
matches
onto an effective current in the chiral lagrangian of the form \cho\
\eqn\current{ \eqalign{
\overline{c}\Gamma b\rightarrow C_{bc} &\
\left( \eta^{(\Lambda)}(w)\overline{T}_c(v^\prime) \Gamma T_b (v) \right. \cr
&\left.   - \left[ g_{\sigma\rho}\eta^{(\Sigma)}_1(w) - v_\sigma v^\prime_\rho
\eta^{(\Sigma)}_2(w) \right]
\overline{S}_c^\sigma (v^\prime) \Gamma S^\rho_b (v)\ \right) + ............}\
\ \ \ \ ,}
where the dots denote terms higher order in $1/m_c$, the light quark mass
matrix
and derivatives.
The Isgur-Wise function $\eta^{(\Lambda)}(w)$ has been defined earlier and
$\eta^{(\Sigma)}_1(w), \eta^{(\Sigma)}_2(w)$ are Isgur-Wise functions for the
weak transition $\Sigma^{(*)}_b \rightarrow \Sigma^{(*)}_c$.
Both $\eta^{(\Lambda)}(w)$ and $\eta^{(\Sigma)}_1(w)$ are normalised to unity
at
zero recoil at which point any dependence on   $\eta^{(\Sigma)}_2(w)$ vanishes.

At leading order in the heavy quark expansion, the heavy quark lagrangian
is independent of the mass of the heavy quark and interactions with the
light degrees of freedom are spin independent.  Consequently, the
$\Sigma^*_Q$ and $\Sigma_Q$ baryons are degenerate.
At next order in $1/m_Q$, the chromomagnetic interaction
\eqn\chromo{ {\cal L}_1 = -Z_Q{1\over 2m_Q} \overline{h}
\sigma_{\alpha\beta} G^{\alpha\beta} h\ \ \ ,}
($G$ is the gluon field strength tensor and $Z_Q$ is a renormalisation
factor) gives rise to a finite mass splitting.
This is reproduced in the chiral lagrangian by a term of the form
\eqn\mass{ {\cal L}_{\Delta m} = {\Delta m\over 6}
\left( g_{\mu\alpha}g_{\nu\beta} - g_{\nu\alpha}g_{\mu\beta}\right)
\overline{S}^\mu i\sigma^{\alpha\beta}S^\nu\ \ \ ,}
where $\Delta m$ is the mass splitting between the $\Sigma^*_Q$ and $\Sigma_Q$
baryons.
In order to make things simple we define two mass differences,
\eqn\diff{ \eqalign{
\Delta_c = &\ m_{\Sigma_c} - m_{\Lambda_c}\cr
\Delta_c^* = &\ m_{\Sigma_c^*} - m_{\Lambda_c}}\ \ \ \ \ ,}
giving $\Delta m = \Delta_c^*-\Delta_c $.
As we are neglecting terms of order $1/m_b$ we also have
\eqn\bbbb{\Delta_0 = \ m_{\Sigma_b} - m_{\Lambda_b} =
m_{\Sigma_b^*} - m_{\Lambda_b} \ \ \ \ .}

The leading long-distance correction to the weak matrix element for \semi\ from
the spin-symmetry breaking
$\Sigma^*_c - \Sigma_c$ mass splitting arises from the graphs shown in
\fig\graphs{Graphs generating the leading long-distance
$1/m_c^n$ corrections to the weak matrix element for
\semi\ that are nonanalytic in the pion mass.}.
These contribute terms that are nonanalytic in the pion mass and are formally
dominant in the chiral limit
(We will only consider contributions from $\pi$'s and not  from $K$'s or
$\eta$'s).
We find that the weak matrix element becomes
\eqn\wkmat{\eqalign{
\langle \Lambda_c (v^\prime) | \overline{c}\Gamma b | &  \Lambda_b(v) \rangle =
C_{bc}\overline{u}_{\Lambda_c}(v^\prime)\Gamma u_{\Lambda_b}(v)
\left(  \eta^{(\Lambda)}(w) \right.\cr
&\left. + {g_3^2\over 4\pi^2f_\pi^2}\left[
\delta \eta^{(\Lambda)}(\Delta_0,\Delta_c^*,\Delta_c,w)
- \delta \eta^{(\Lambda)}(\Delta_0,\Delta_0,\Delta_0,w)\right] \  \right)} \ \
\ \ ,}
where
\eqn\deleta{\eqalign{
\delta \eta^{(\Lambda)}(\Delta_0,\Delta_c^*,\Delta_c,w) = &
\eta^{(\Sigma)}_1(w) \left[
(w^2+2)[I_1(\Delta_0,\Delta_c^*,w) + {1\over 2}I_1(\Delta_0,\Delta_c,w)]
\right.\cr
&\left.\phantom{aaaa{1\over 2}}
-w(w^2-1)[I_2(\Delta_0,\Delta_c^*,w) + I_2(\Delta_0,\Delta_c,w)] \right]  \cr
-\eta^{(\Sigma)}_2(w)  ( w^2 & -1) \left[
w [I_1(\Delta_0,\Delta_c^*,w) + {1\over 2}I_1(\Delta_0,\Delta_c,w)] \right.\cr
&\left.\phantom{aaa{1\over 2}}
-(w^2-1)[ I_2(\Delta_0,\Delta_c^*,w) + {1\over 2}I_2(\Delta_0,\Delta_c,w) ]
\right] \cr
  -{3\over 4}\eta^{(\Lambda)}(w) &\left[
3I_1(\Delta_0,\Delta_0,1) + 2I_1(\Delta_c^*,\Delta_c^*,1)
+I_1(\Delta_c,\Delta_c,1) \right]    }\ \ \ \ \ .}
The contribution proportional to $\eta^{(\Lambda)}(w)$ comes from
wavefunction renormalisation while the remaining terms
come from vertex graphs shown in \graphs .
The integral $I_1$ is
\eqn\inta{\eqalign{
I_1(\Delta_1, & \Delta_2,w)   =  \int_0^1 dx\ \left[
-\left({b(x)\over a(x)}\right)^2\log\left(m_\pi^2/\Lambda_\chi^2\right)
+{b(x)\pi m_\pi\over ( \sqrt{a(x)}) ^3}  \right.\cr
&\left.
+ {b(x)m_\pi\over ( \sqrt{a(x)}) ^3}\left(
\left(\sqrt{{b(x)^2\over m_\pi^2 a(x)} - 1}\right) \log\left(\lambda (x)\right)
-\pi + 2{b(x)\over m_\pi \sqrt{a(x)}}\right) \right]
}\ \ \ \ \ ,}
where $m_\pi$ is the pion mass and
\eqn\funs{\eqalign{
\lambda(x) = &\ {{b(x)\over m_\pi \sqrt{a(x)}}-\sqrt{{b(x)^2\over m_\pi^2 a(x)}
- 1+i\epsilon}\over
{b(x)\over m_\pi \sqrt{a(x)}}+\sqrt{{b(x)^2\over m_\pi^2 a(x)} - 1+i\epsilon}}
 \cr
a(x) = & \ 1+2x(1-x)(w-1) \cr
b(x) = & \ x\Delta_1 + (1-x)\Delta_2 }\ \ \ .}
The integral $I_2$ is
\eqn\intb{
I_2(\Delta_1,\Delta_2,w) = 4\int_0^1 dx\ x(1-x) {\partial\over\partial
a(x)}\left[ ........\right] \ \ \ \ ,}
where the dots represent the function inside the square bracket of $I_1$.

It is easy to show that $\delta \eta^{(\Lambda)}(\Delta_0,\Delta_0,\Delta_0,w)$
vanishes at zero recoil, $w=1$ as expected, as this corresponds to the
$m_c\rightarrow\infty$ limit.
This function can be absorbed by a redefinition of $\eta^{(\Lambda)}(w)$.
It is $\delta \eta^{(\Lambda)}(\Delta_0,\Delta_c^*,\Delta_c,w)$ that contains
the
$1/m_c^{2+n} \ \ n=0,1,2,.....$ corrections nonanalytic in the pion mass that
do not vanish at zero recoil and also terms $1/m_c^{1+n} \ \ n=0,1,2,.....$
that
are nonzero away from zero recoil.
It is important to observe that these long-distance effects only contribute to
the
$\gamma^\mu$ and $\gamma^\mu\gamma_5$ formfactors, and not to the
$v^\mu$ and $v^\mu\gamma_5$ formfactors ($f_2, f_3,g_2$ and $g_3$) .
Such formfactors are allowed by the b quark spin symmetry, and would
correspond to a contribution to  ${\cal F}_2$ in \inter\ .

We have chosen the chiral symmetry breaking scale $\Lambda_\chi$ as the
renormalisation point at which  to evaluate
\inta\ and \intb\ .
The scale dependence of these graphs is compensated by an equal and
opposite scale dependence of a local counterterm of the form
\eqn\counter{{\cal L}_{\rm ct} = C(\mu)
\overline{T_c}\sigma^{\alpha\beta}{1\over 2}
(1+\slash{v})
\sigma^{\sigma\rho}{1\over 2}(1+\slash{v})\Gamma T_b K_{\alpha\beta\sigma\rho}\
\ \ ,}
where $K_{\alpha\beta\sigma\rho}$ is some function of $v$, $v^\prime$ and
$g^{\lambda\gamma}$ and is pairwise antisymmetric on $\alpha\beta$ and
$\sigma\rho$.
This is similar to the construction of \randwise\chowwise\ .
It corresponds to two insertions of the chromomagnetic operator on the
charm quark line, and is formally subdominant compared to the nonanalytic terms
found from \deleta\ .   It is the scale dependence of the coefficient $C(\mu)$
that
compensates the scale dependence of the logarithms in  \inta\ and \intb\ .

At the zero recoil point $w=1$ we find that the contribution from
$\delta\eta^{(\Lambda)}(\Delta_0,\Delta_c^*,\Delta_c,1)$  linear in $\Delta m$
vanishes, giving a leading contribution of order $(\Delta m)^2$ as required.
Neither the $\Sigma_c^*$ or  $\Sigma_b^{(*)}$ baryons have been observed yet
and
in order to get an estimate of the size of corrections at the zero recoil point
we will
use a  nonrelativistic quark model calculation of the $\Sigma_c^*$ mass
$m_{\Sigma_c^*} = 2494\pm 16$MeV
\ref\paul{R. E. Cutkosky and P. Geiger,\physrev{D48}{1993}{1315}.}.
This value, combined with the experimental measurements of the other relevant
masses
\ref\pdg{Particle Data Group, \physrev{D45}{1992}{1}.}
gives
\eqn\masses{
\Delta_c^* = \ 209\pm 16 {\rm MeV} \ \ \ ,\ \ \
\Delta_c = \ 167.8 \pm 0.4 {\rm MeV}  \ \ \ \ ,}
and hence $\Delta m = 41\pm 16 {\rm MeV}$,
much smaller than the corresponding $D^*-D$ mass splitting in the meson sector
\foot{For other estimates of $\Delta m$ see
\ref\KROQ{W. Kwong, J. Rosner and C. Quigg, Ann. Rev. Nucl. Sci., 37 (1987)
325.}
\ref\fp{A.F. Falk and M.E. Peskin, SLAC-PUB-6311 (1993).}
and for the present experimental situation see
\ref\argus{H. Albrecht etal., (ARGUS), \pl{211}{1988}{489}.}\nref\cleo{T.
Bowcock etal.,
(CLEO), \prl{62}{1989}{1240}.}-\ref\tpc{J.C. Anjos etal., (Tagged Photon
Collaboration),
\prl{62}{1989}{1721}.}.} .
We use this to estimate an intrinsic mass splitting of
$\Delta_0=194\pm 11 {\rm MeV}$ where
the uncertainty depends entirely on that of $m_{\Sigma_c^*}$.
Using the above baryon masses and uncertainties we find that the correction
at zero recoil is
\eqn\zero{
\delta\eta^{(\Lambda)}(\Delta_0,\Delta_c^*,\Delta_c,1)
= -(3.4 \pm 2.0) \times 10^{-4} g_3^2 \ \ \ \ .}
The axial coupling constant $g_3$ is, as yet,  undetermined.
However, in the large-$N_c$  limit of QCD ( $N_c$ is the number of colours)
it has been shown to be related to the $\pi$-N axial coupling constant
$g_3=\sqrt{3\over 2}g_A$ where $g_A=1.25$
\ref\gural{Z. Guralnik, M. Luke and A.V. Manohar, \np{390}{1993}{474}.}
\ref\jenkN{E. Jenkins, \pl{315}{1993}{431}.} .
For the purposes of this work we will assume that $g_3$ is of order unity.
This correction is much smaller than the corresponding correction found
in the meson sector \randwise\ for two reasons.
Firstly, the spin dependent mass splitting is much smaller in the baryon sector
than in
the meson sector.
Secondly, there is a cancellation between the $\log m_\pi^2$ term (which
contributes
$ - (1.8 \pm 1.2)\times 10^{-3}$) and the
remaining terms (which depend strongly on $\Delta_0$).
The corrections are suppressed by the intrinsic
$\Sigma_Q^{(*)}-\Lambda_Q$ mass splitting, which has no analogue in the meson
sector.

We can interpret this result in a variety of ways.
The most favorable interpretation would be that indeed the leading
long-distance
$1/m_c^{2+n},\ \ n=0,1,...$
corrections at  zero recoil are small.  However, it is more likely that
in fact these graphs are not dominant over the
local counter terms due to cancellations arising from the intrinsic
$\Sigma_Q^{(*)}-\Lambda_Q$ mass splitting.
In this case the zero recoil corrections are incalculable within the framework
of chiral perturbation theory.
As  has been stressed throughout this work the modifications we
have computed arise from the spin symmetry breaking
$\Sigma_Q^*-\Sigma_Q$ mass splitting.
Corrections to the zero recoil point may also arise from the operator
\eqn\fermi{{\cal L}_1 = {1\over 2m_Q}\overline{h}(iD)^2h\ \ \ ,}
in the heavy quark lagrangian
corresponding to the fermi motion of the heavy quark within the baryon.
However, they are not computable
within chiral perturbation theory and would appear as local counterterms at
order $1/m_c^2$ with unknown coefficients.
An estimate of such corrections has been made in the large-$N_c$ limit of QCD
\ref\jmw{E. Jenkins, A.V. Manohar and M.B. Wise, \np{396}{1993}{38}.}
where a baryon containing a heavy quark appears as a bound state
of a light baryon skyrmion and a heavy meson matter field.
Zero recoil corrections arise from the fermi motion of both the nucleon
and heavy meson forming the bound state
and are found  to be about $1\%$, much larger than the
those computed in this work.

There are analogous corrections to the weak decays
$\Xi_{b1}\rightarrow\Xi_{c1}e^-\overline{\nu}_e$, arising from the spin
symmetry breaking
$\Xi_{c2}^*-\Xi_{c2}$ mass splitting, $\Delta m_{\Xi_c}$.
We find the corrections are $1/4$ the size of the corrections for  \semi\
using
quark model estimates for $\Delta m_{\Xi_c}$.
This reduction is mainly a result of the size of the relevant Clebsch-Gordan
coefficients.

In conclusion,
we have computed the formally dominant corrections, nonanalytic in the pion
mass,
to the \semi\ matrix element arising from the
spin symmetry breaking $\Sigma_c^*-\Sigma_c$ mass splitting, summing a class of
$1/m_c^{2+n}\ \ n=0,1,2,......$ terms  that contribute at zero recoil and a
class of
$1/m_c^{1+n}\ \ n=0,1,2,......$ terms  that contribute away from zero recoil.
We find that these corrections do not induce the lorentz structures $v^\mu,
v^\mu\gamma_5$
even though such forms are allowed by the $b$-quark spin symmetry.
The contributions to the formfactors of $\gamma^\mu, \gamma^\mu\gamma_5$ are
equal
($b$-quark spin symmetry) and are found to be
$ - (3.4 \pm 2.0) \times 10^{-4} g_3^2$ at the zero recoil point.
Computations in the large-$N_c$ limit of QCD indicate corrections
arising from the fermi-motion of the heavy quark inside the hadron may be
substantially
larger than this.
Such contributions will only appear as local counterterms in the chiral
lagrangian.
Hence it is possible and quite likely that local counterterms give corrections
much larger than
those induced by the spin symmetry breaking mass term as computed in this work.

\bigskip

I would like to thank M.B.Wise, A. Falk and M. Luke  for useful discussions.
This research was supported in part by
the Department of Energy under contract DE--FG02--91ER40682.

\listrefs
\listfigs
\vfill\eject

\insertfig{Figure 1 }{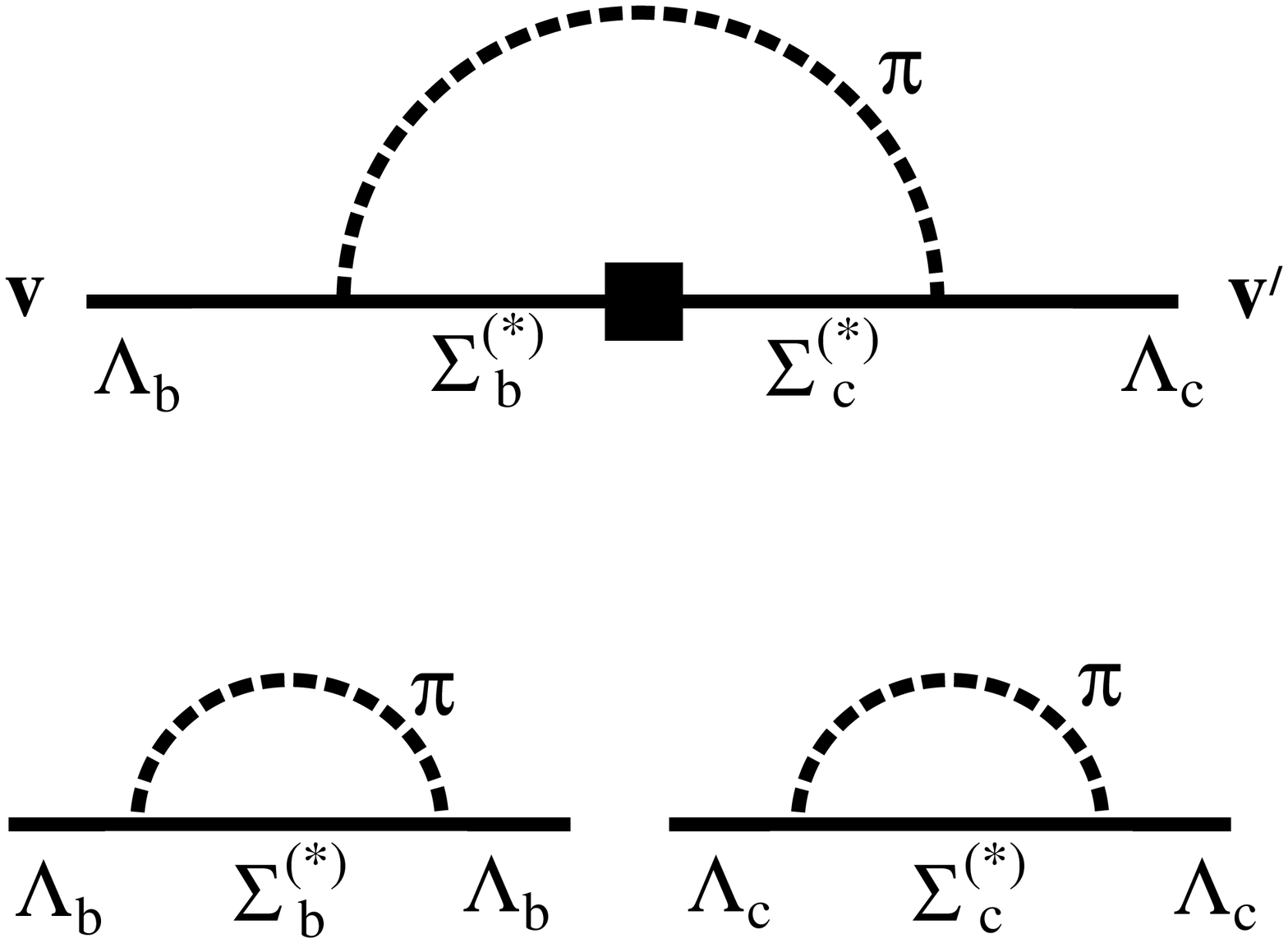}

\vfill\eject
\bye